\documentclass[twocolumn,showpacs,preprintnumbers,amsmath,amssymb,prl]{revtex4}

\usepackage{graphicx}
\usepackage{dcolumn}
\usepackage{bm}
\usepackage{multirow}
\usepackage{color}

\def\sub#1{_{\mathrm{#1}}}

\def\Vec#1{\boldsymbol #1}

\def\3He{\mbox{$^3$He}}
\def\4He{\mbox{$^4$He}}

\begin{document}

\preprint{}

\title{Collision Dynamics and Rung Formation of Non-Abelian Vortices}

\author{Michikazu Kobayashi$^1$, Yuki Kawaguchi$^1$, Muneto Nitta$^2$, and Masahito Ueda$^{1,3}$}

\affiliation{$^1$Department of Physics, University of Tokyo, Hongo 7-3-1, Bunkyo-ku, Tokyo 113-0033, Japan.}

\affiliation{$^2$Department of Physics, Keio University, Hiyoshi, Yokohama, Kanagawa 223-8511, Japan}

\affiliation{$^3$ERATO Macroscopic Quantum Control Project, JST, Tokyo 113-8656, Japan.}

\date{\today}

\begin{abstract}
We investigate the collision dynamics of two non-Abelian vortices and find that, unlike Abelian vortices, they neither reconnect themselves nor pass through each other, but create a rung between them in a topologically stable manner.
Our predictions are verified using the model of the cyclic phase of a spin-2 spinor Bose-Einstein condensate.
\end{abstract}


\maketitle

Quantized vortices are topological defects in the superfluid order-parameter, and their character depends on the topology of the order-parameter manifold of the system.
In the case of single component Bose-Einstein Condensates (BECs), for example, the order-parameter manifold is $U(1)$ and the quantized vortices are characterized by an additive group of integers.
However, the situation is dramatically different for spinor BECs, where some phases accommodate non-Abelian vortices where the collision dynamics shows markedly different behavior from that of Abelian vortices.
In this Letter, we investigate the collisional properties of non-Abelian vortices and their unique topological properties.
In particular, we find that a rung vortex that bridges the colliding vortices is always formed upon collision, regardless of the ranges of the kinematic parameters.
We verify these ideas with numerical simulations for a spin-2 BEC.

The topological charge of a vortex can be identified by studying how the order-parameter changes along a closed path encircling the vortex.
For $U(1)$ vortices, the $U(1)$ phase changes around the vortex by an integer multiple of $2 \pi$, and the topological charge is expressed by integers.
In spinor BECs, on the other hand, the change of the order-parameter around the vortex involves not only the $U(1)$ phase but also an $SO(3)$ rotation of the spin.
Therefore, there are phases in which topological charges of vortices do not commute with each other and form a non-Abelian group.
We define such vortices with noncommutative topological charges as non-Abelian vortices \cite{definition}.

A salient feature of non-Abelian vortices manifests itself in the collision dynamics.
In Abelian vortices, the following three types of collisions are possible: reconnection, passing through, and a formation of a rung vortex that bridges the colliding vortices.
The reconnection of Abelian $U(1)$ vortices has been studied theoretically \cite{Koplik}, and observed recently in superfluid \4He \cite{Bewley}.
Moreover, there are some theoretical works which predict the rung structure that connects two attracting $U(1)$ vortices \cite{Laguna} or $U(1) \times U(1)$ vortices \cite{Bevis}.

\begin{figure}[tbh]
\centering
\includegraphics[width=0.95\linewidth]{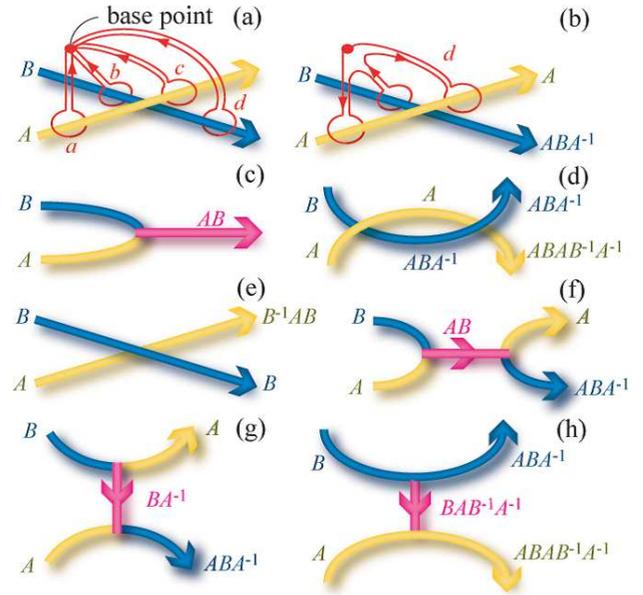}
\caption{\label{fig-algebra} (Color) (a) Four closed paths for two colliding vortices, where $A$ and $B$ denote the corresponding topological charges. (b) Path homotopic to path $d$ in (a), and topological charges of vortices. (c) Y-shaped junction. (d) Linked vortices. (e)-(g) Collision patterns from two vortices starting from the configuration in (a). (e) Passing through. (f) Rung $AB$ between two vortices. (g) Rung $BA^{-1}$ between two vortices. (h) Collision of linked vortices and rung $BAB^{-1}A^{-1}$.}
\end{figure}
When the vortices are non-Abelian, the situation changes dramatically.
It was predicted that the collision of two non-Abelian vortices produces a tangled and connected rung vortex \cite{Mermin}.
In fact, for the case of two vortices with noncommutative topological charges, reconnection and passing through are topologically forbidden and only the formation of a rung vortex is allowed.
This can be understood by considering algebraic and geometric structures of the vortices as shown in Fig \ref{fig-algebra}.
Let us consider two colliding vortices with topological charges $A$ and $B$ as illustrated in Fig. \ref{fig-algebra} (a).
The topological charge of a vortex can be determined from a change of the order parameter along a closed path that encircles the vortex.
We consider two colliding vortices and four paths $a$, $b$, $c$, and $d$, as shown in Fig. \ref{fig-algebra} (a), and assume that paths $a$ and $b$ define the topological charges of the vortices as $A$ and $B$, respectively.
When the base point is fixed, paths $a$ and $c$ are topologically equivalent.
On the other hand, path $d$ is topologically different from path $b$, but it can be continuously deformed to the path shown in Fig. \ref{fig-algebra} (b).
As a result, path $d$ defines the topological charge as $ABA^{-1}$.
When the charges $A$ and $B$ are noncommutative for non-Abelian vortices, paths $b$ and $d$ give different topological charges for the same vortex.
Similarly, we can define the topological charges of vortices for a Y-shaped junction and for linked vortices as shown in Figs. \ref{fig-algebra} (c) and \ref{fig-algebra} (d), respectively.
Upon the collision of two vortices [Fig. \ref{fig-algebra} (b)], three possible consequences [Figs. \ref{fig-algebra}(e) - \ref{fig-algebra} (g)] follow, where Fig. \ref{fig-algebra} (e) shows ``passing through", and Figs. \ref{fig-algebra} (f) and \ref{fig-algebra} (g) show formation of ``rungs" that bridges the two vortices.
When $A$ and $B$ are commutative, all three cases are topologically allowed.
If $A$ and $B$ are the same topological charge, in particular, Fig. \ref{fig-algebra} (g) reduces to reconnection because the rung vanishes identically ($BA^{-1} = 1$).
When $A$ and $B$ are noncommutative for non-Abelian vortices, however, the transition from Figs. \ref{fig-algebra} (b) - \ref{fig-algebra} (e) is topologically forbidden because they are topologically distinct.
Therefore, a rung with the topological charge of $AB$ or $BA^{-1}$ must be formed after the collision, regardless of the kinematic parameters such as the collision angles and the initial relative speed.
From the viewpoint of the stability of rung formation, this presents a great contrast to that of Abelian vortices, because in the case of Abelian vortices, the formation of a rung is usually energetically unfavorable and occurs only in specific situations such as the collision of attractive vortices as in type-I superconductors \cite{Laguna}, and whether a rung is formed or not is strongly dependent on the kinematic parameters of the collision \cite{Laguna,Bevis}.
The other important property of non-Abelian vortices arises from the collision of linked vortices as shown in Figs. \ref{fig-algebra} (d) and \ref{fig-algebra} (h).
The rung formed from the linked vortices shown in Fig. \ref{fig-algebra} (h) arises from the genuine non-Abelian character because the rung $BAB^{-1}A^{-1}$ always vanishes for Abelian vortices.
Therefore, linked vortices with commutative topological charges can unravel, whereas linked non-Abelian vortices with noncommutative ones cannot.

The novel feature of the collision dynamics has been theoretically investigated in several systems such as a network of cosmic strings in the Universe and disclinations in biaxial nematic liquid crystals.
In the context of cosmic strings, collisions of vortices also play an important role, because it determines the number density of cosmic strings in the Universe.
Abelian cosmic strings can reconnect \cite{VS,cosmic_strings} and decrease their number by themselves \cite{VS}, consistent with the current measurements of the cosmic microwave background \cite{observation}.
In non-Abelian cosmic strings, on the other hand, reconnection does not occur, but robust rungs are formed which develop into a junction or a network structure \cite{Spergel:1996ai,Bucher:1998mh}.
Similar properties of non-Abelian vortices have been studied for disclinations in biaxial nematic liquid crystals \cite{de-Gennes,Zapotocky}, the topological charge of which is expected to belong to the non-Abelian quaternion group $Q_8$.
It has been reported \cite{Zapotocky} that there is a qualitative difference in coarsening dynamics between Abelian and non-Abelian disclinations, and the latter dynamics is slower than the former due to formations of rungs among disclinations.

In this Letter, we propose a spinor BEC as an ideal system to study non-Abelian vortices and their dynamics.
The major advantages of this system are that the microscopic Hamiltonian is known and that the dynamics of the system can be investigated in real time. 
Below we demonstrate that non-Abelian vortices appear in the cyclic phase of a spin-2 spinor BEC and show that the collision dynamics obeys the aforementioned algebraic rules.
Recently, spin-2 BECs have been investigated in $F = 2$ $^{87}$Rb atoms \cite{Schmaljohann} and various features have been predicted \cite{Koashi,Ciobanu,Makela,Saito,Pogosov,Zhou-2,Semenoff,Barnett}.
Although the ground-state phase of an $F = 2$ $^{87}$Rb BEC is widely believed to be antiferromagnetic, the possibility of the cyclic phase has not yet been excluded due to complications arising from quadratic Zeeman effects and hyperfine-spin-exchanging relaxations \cite{Tojo}.
Topological charges of vortices in the cyclic phase of the spin-2 BEC are expressed by the discrete tetrahedral non-Abelian group \cite{Makela,Semenoff}.
We will show below that the stable rung can be formed resulting from the algebra shown in Fig. \ref{fig-algebra}.

We start with a BEC of spin-2 atoms with mass $M$ whose energy functional is given by \cite{Koashi,Ciobanu}
\begin{align}
H = \int d^3 r \: \Big[ & \sum_{m = -2}^{2} \Psi_m^\ast \Big( -\frac{\hbar^2}{2M} \nabla^2 \Big) \Psi_m \nonumber \\
& + \frac{c_0}{2} n\sub{tot}^2 + \frac{c_1}{2} |\Vec{F}|^2 + \frac{c_2}{2} |A_{00}|^2 \Big], \label{eq-Hamiltonian}
\end{align}
where $\Psi_m$ is the order-parameter of the BEC in a magnetic sublevel $m = 0, \pm 1, \pm 2$ at position $\Vec{r}$, and $n\sub{tot} = \sum_{m = -2}^{2} |\Psi_m|^2$, $\Vec{F} = \sum_{m, m^\prime = -2}^{2} \Psi_m^\ast \Vec{F}_{m, m^\prime} \Psi_{m^\prime}$, and $A_{00} = (2 \Psi_2 \Psi_{-2} - 2 \Psi_1 \Psi_{-1} + \Psi_0^2) / \sqrt{5}$ are the total number density, the spin vector density, and the spin-singlet pair amplitude, respectively.
Here $\Vec{F}_{m,m^\prime}$ is a vector of spin-2 matrices.
The ground state of the cyclic phase is realized for $c_1 > 0$ and $c_2 > 0$, where both $\Vec{F}$ and $A_{00}$ vanish.
One representative state for the cyclic phase is given by $\Psi\sub{cyclic} = \sqrt{n\sub{tot}} (i/2, 0, 1/\sqrt{2}, 0, i/2)^T$.

Through the $U(1)$ gauge transformation and the $SO(3)$ spin rotation, it is possible to transform from one to another cyclic state, $\Psi^\prime\sub{cyclic} = e^{i \phi} e^{-i \Vec{F} \cdot \hat{\Vec{\omega}} \theta} \Psi\sub{cyclic}$, where $\phi$ is the $U(1)$ gauge, and $\hat{\Vec{\omega}}$ and $\theta$ are the unit vector of the rotational axis and angle of the spin rotation, respectively.
In the cyclic phase, $\Psi\sub{cyclic}$ is invariant under the following 12 transformations: $\Vec{1}$, $I_x = e^{i F_x \pi}$, $I_y = e^{i F_y \pi}$, $I_z = e^{i F_z \pi}$, $\bar{C} = e^{2 \pi i / 3} e^{- 2 \pi i (F_x + F_y + F_z) / 3 \sqrt{3}}$, $\bar{C}^2$, $I_x \bar{C}$, $I_y \bar{C}$, $I_z \bar{C}$, $I_x \bar{C}^2$, $I_y \bar{C}^2$, and $I_z \bar{C}^2$ \cite{Semenoff}.
The overbar is added to emphasize that the operation includes not only a rotation in spin space but also a gauge transformation.
These 12 transformations form the non-Abelian tetrahedral group $T$. 

Topological charges of vortices can be classified by 12 elements of the non-Abelian group $T$.
We represent these topological charges as $\Vec{1}$, $I_x$, $I_y$, $\cdots$.
Vortices are also classified into four conjugacy classes : (I): integer vortex; $\Vec{1}$, (II): 1/2 - spin vortex; $I_x$, $I_y$, and $I_z$, (III): 1/3 vortex; $\bar{C}$, $I_x \bar{C}$, $I_y \bar{C}$, and $I_z \bar{C}$, and (IV): 2/3 vortex; $\bar{C}^2$, $I_x \bar{C}^2$, $I_y \bar{C}^2$, and $I_z \bar{C}^2$.
Topological charges in the same conjugacy class transform into one another under the global gauge transformation and the spin rotation.
The order parameters for straight vortices along the $z$-axis in each conjugacy class can be written in cylindrical coordinates $(r, \varphi, z)$ as
\begin{align}
\Psi & = \frac{1}{2} \sqrt{n\sub{tot}(r)} e^{i n_1 \varphi} \hat{S} \nonumber \\
& \times \left\{ \begin{array}{l}
\displaystyle \left( i f e^{2 i (n_2 + 1) \varphi}, 0, \sqrt{2} h, 0, i f e^{- 2 i (n_2 + 1) \varphi} \right)^{T} \\
\displaystyle \left( i f e^{i (2 n_2 + 1) \varphi}, 0, \sqrt{2} h, 0, i f e^{- i (2 n_2 + 1) \varphi} \right)^{T} \\
\displaystyle \frac{2}{\sqrt{3}} \left( f e^{i (2 n_2 + 1) \varphi}, 0, 0, \sqrt{2} g e^{- i n_2 \varphi}, 0 \right)^{T} \\
\displaystyle \frac{2}{\sqrt{3}} \left( f e^{i (2 n_2 - 1) \varphi}, 0, 0, \sqrt{2} g e^{- i n_2 \varphi}, 0 \right)^{T}
\end{array} \right. \label{eq-vortex-wave-function}
\end{align}
for (I), (II), (III), and (IV), respectively, where the vortex is placed at $r = 0$.
Here, $n_1$ and $n_2$ are the integer winding numbers, $f = f(r)$, $g = g(r)$, and $h = h(r)$ are real functions that satisfy $[f(r)^2 + h(r)^2] / 2 = [f(r)^2 + 2 g(r)^2] / 3 = 1$ and $f(r \to \infty) = g(r \to \infty) = h(r \to \infty) = 1$; $\hat{S}$ represents the arbitrary global gauge transformation and global spin rotation.
At the vortex core, the cyclic order parameter changes to that of a different phase.
With the minimum windings ($n_1 = n_2 = 0$), we obtain the core structure of each conjugacy class by taking $f(r = 0) = 0$ as: (I)(II) $\Psi \propto \hat{S} ( 0, 0, 1, 0, 0 )^{T}$ and (III)(IV) $\Psi \propto \hat{S} ( 0, 0, 0, 1, 0 )^{T}$, {\it i.e.}, the core of (I) and (II) vortices have a finite spin-singlet pair amplitude ($A_{00} \neq 0$), and that of (III) and (IV) vortices have a finite magnetization ($\Vec{F} \neq 0$) \cite{core}.

\begin{figure}[tbh]
\centering
\includegraphics[width=0.9\linewidth]{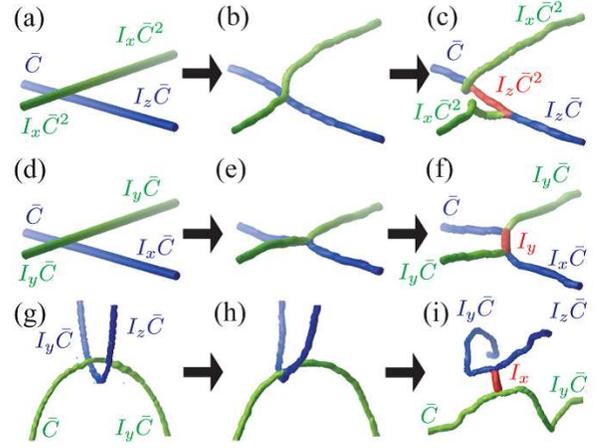}
\caption{\label{fig-simulation} (Color) Collision dynamics of two non-Abelian vortices \cite{movie}. (a)-(c): Formation of an $\Vec{F} \neq 0$ rung from two straight vortices. (d)-(f): Formation of an $A_{00} \neq 0$ rung from two straight vortices. (g)-(i): Formation of an $A_{00} \neq 0$ rung from two linked vortices. In all the figures, the isosurfaces of magnetization $|\Vec{F}|^2$ and the singlet-pair amplitude $|A_{00}|^2$ are shown for vortex cores with $|\Vec{F}|^2 \neq 0$ and $|A_{00}|^2 \neq 0$, respectively [see Eq. (\ref{eq-vortex-wave-function}) and the following sentences]. The topological charge of each vortex is also indicated.}
\end{figure}
When there is more than one vortex in the system, each topological charge cannot independently transform into one another under the global gauge transformation and the global spin rotation.
Therefore, the relative relationship of their topological charges (commutative or noncommutative, in particular) becomes important and their collision dynamics becomes nontrivial.
To investigate the detailed dynamics of vortex collisions, we numerically solve the nonlinear Schr\"odinger equation \cite{Saito,Pogosov} derived from Eq. (\ref{eq-Hamiltonian}) in a uniform box subject to the Neumann boundary condition, starting from the two types of initial conditions as shown in Fig. \ref{fig-simulation}: (I) two straight vortices at an oblique angle [Fig. \ref{fig-simulation} (a) and \ref{fig-simulation} (d)], and (II) two linked vortices [Fig. \ref{fig-simulation}(g)].
We take $c_1 = c_2 = 0.5 c_0$, for which the cyclic phase and non-Abelian vortices discussed above can exist stably.
In the present simulation, we perform the collision of vortices with $\Vec{F} \neq 0$ cores and topological charges shown in Figs. \ref{fig-simulation} (a), \ref{fig-simulation} (d), and \ref{fig-simulation} (g).
After the collision, two vortices get connected and a rung appears between the two vortices.
Depending on the initial topological charges of the vortices, we obtain rungs with $\Vec{F} \neq 0$ core [Fig. \ref{fig-simulation} (c)] or $A_{00} \neq 0$ core [Figs. \ref{fig-simulation} (f) and \ref{fig-simulation} (i)].
For the collisions of straight vortices, topological charges of rungs in Figs. \ref{fig-simulation} (c) and \ref{fig-simulation} (f) obey the algebras shown in Fig. \ref{fig-algebra} (g), namely $\bar{C} \: (I_x \bar{C}^2)^{-1} = I_z \bar{C}^2$ and $\bar{C} \: (I_y \bar{C})^{-1} = I_y$, respectively.
We have performed numerical simulations with various combinations of topological charges, relative velocities, and collision angles, and confirmed that passing through and reconnection occur only when the topological charges of the two vortices are commutative, and that the formation of a rung always occurs, when the topological charges of the two vortices are noncommutative.
For the linked vortices shown in Fig. \ref{fig-simulation} (g), we can expect the formation of a rung as shown in Fig. \ref{fig-algebra} (h).
The formed rung in Fig. \ref{fig-simulation} (i) satisfies expected algebra: $I_y \bar{C} \: \bar{C} \: (I_y \bar{C})^{-1} \: (\bar{C})^{-1} = I_x$.
We also have checked that unraveling of two linked vortices never happens for noncommutative topological charges.

We finally describe a possible experimental manifestation of rungs.
The phase-contrast imaging experiment \cite{Higbie} enables the measurement of local magnetization, and vortices with $\Vec{F} \neq 0$ cores appear as localized magnetization lines.
For example, rungs with $\Vec{F} \neq 0$ cores like Fig. \ref{fig-simulation} (c) manifest themselves as bridged structures of localized magnetization.

In conclusion, we have algebraically studied the collision dynamics of non-Abelian vortices.
After the collision, two non-Abelian vortices with noncommutative topological charges neither reconnect themselves nor pass through each other, but always create a rung between them.
We also have substantiated our theory by carrying out numerical simulations for vortices classified by the discrete tetrahedral non-Abelian group in the cyclic phase of a spin-2 spinor BEC.
We expect that the results of our study may find applications in other non-Abelian systems such as cosmic strings, biaxial nematic liquid crystals, and superconductors with high internal degrees of freedom.
In particular, in quantum turbulence, where the key process is the reconnection of quantized vortices \cite{Niemela}, we expect that a dramatic change should occur with non-Abelian vortices, as reported in other systems \cite{Spergel:1996ai,Zapotocky}, and that the non-Abelian properties found in this Letter will open up a new research field of non-Abelian quantum turbulence.

This work was supported by MEXT, Japan (KAKENHI, No. 207229, 17071005, and 20740141, the Global COE Program ``the Physical Sciences Frontier", and the Photon Frontier Network Program).


\begin{thebibliography}{99}
\bibitem{definition}
The topological charges of quantized vortices are defined by the first homotopy group $\pi_1$ of the order-parameter manifold $G/H$ of the system, where $G$ is the symmetry of the Hamiltonian and $H$ is the remaining symmetry of the order-parameter.
In general, three different definitions can be adopted for non-Abelian vortices; (i) $G$ is non-Abelian, (ii) $H$ is non-Abelian, and (iii) $\pi_1(G/H)$ is non-Abelian.
Although definition (ii) is widely used, in this Letter, we adopt definition (iii).
\end{thebibliography}
\end{document}